\documentclass[11pt,a4paper]{article}
\usepackage{amsmath,amssymb}
\usepackage{epsfig,graphicx}

\textheight
24.6cm
\textwidth
15.9cm

\begin{document}

\title{\bf The \boldmath $Z\to c\bar c\to\gamma\gamma^{\ast}$, $Z\to b\bar b\to\gamma\gamma^{\ast}$ triangle diagrams and the
$Z\to\gamma\psi$, $Z\to\gamma\Upsilon$ decays}
\author{N.~N.~Achasov
\footnote{{\bf e-mail}: achasov@math.nsc.ru}
\\
 \small{\em S.L. Sobolev Institute for Mathematics} \\ \small{\em
Academician Koptiug Prospekt, 4,
  Novosibirsk, 630090, Russia }}
\date{}
\maketitle

\begin{abstract}
It is expounded  the approach to the $Z\to\gamma\psi$ and
$Z\to\gamma\Upsilon$ decay study, based on the sum rules for the
$Z\to c\bar c\to\gamma\gamma^{\ast}$ and $Z\to b\bar b\to\gamma
\gamma^{\ast}$ amplitudes and their derivatives.
 The branching ratios of the $Z\to\gamma\psi$ and
$Z\to\gamma\Upsilon$ decays are calculated for different guesses
as to saturation of the sum rules. The lower bounds of
$\sum_{\psi} BR(Z\to\gamma\psi) =  1.95\cdot 10^{-7}$ and
$\sum_{\Upsilon} BR(Z\to\gamma\Upsilon) =  7.23\cdot 10^{-7}$ are
found. Deviations from the lower bounds are discussed, among them
the possibility of $BR(Z\to\gamma J/\psi(1S))\sim
BR(Z\to\gamma\Upsilon(1S))\sim 10^{-6}$, that could be probably
measured in LHC. The angle distributions in the $Z\to\gamma\psi$
and $Z\to\gamma \Upsilon$ decays are calculated also.
\end{abstract}

\section{Introduction}
   The purpose of this paper is to revise  the study of the $Z\to\gamma\psi$ and
$Z\to\gamma \Upsilon$ decays in a dispersion approach
\cite{achasov-Z91,achasov-Z92}
\footnote{Note that   an analogous
dispersion approach was used  to investigate the decays involved
the Higgs boson: $H\rightarrow \gamma \psi\,, \gamma \Upsilon$ and
of the decays $\psi\,, \Upsilon \rightarrow \gamma H$(or axion)
\cite{besprozvannykh-91}.}.

\begin{figure}
\includegraphics[width=0.8\textwidth,height=5cm]{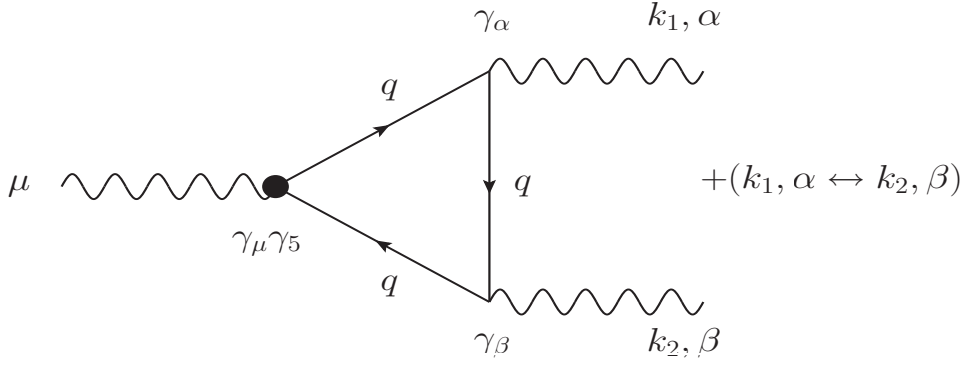}
\caption{The triangle diagrams } \label{fig}
\end{figure}

 Section 2
is devoted to the $Z\to\gamma\psi$ and $Z\to\gamma\Upsilon$
decays.
  Here the invariant amplitudes of the
triangle loop diagrams ,  Figure \ref{fig}, describing the
transition of the axial-vector current $\to q\bar q\to\gamma(k_1)
\gamma(k_2)$ at $k_1^2=0$ and $k_2^2\not=0$, \footnote{These
amplitudes  are calculated in Ref. \cite{achasov-Z92} and can be
found also in Refs. \cite{achasov-AxialAnomaly,achasov-Pole93}.
These calculations made it possible to show
\cite{achasov-Pole93,achasov-Pole92} that in the chiral limit
there is the massless particle-like pole in the transverse part of
the axial-vector channel of the $axial-vector\ current$ $\to q\bar
q \to$ $vector\ current$ $(k_1)$ $\times$ $vector\ current$
$(k_2)$ amplitude at $k_1^2=0$ and $k_2^2\ne0$ in parallel with
the massless particle-like pole in the longitudinal one, generally
accepted at that time.} are used to construct the sum rules for
the    $Z\to c\bar c\ (\mbox{or}\ b\bar b)\to\gamma\gamma\ast$
amplitude and its derivative. Then  the minima of the
 branching ratio sums ($\min\sum_VBR(Z\to\gamma V)$, where $V=\psi$ or $\Upsilon$)
  are  evaluated for different guesses
as to saturation of the sum rules. Three assumption are
investigated: i)   the resonance saturation of the sum rule for
the amplitude in Section 2.1, ii) the resonance saturation of the
sum rule for the amplitude derivative in Section 2.2, iii) and the
simultaneous resonance saturation of the amplitude and its
derivative in Section 2.3. In Sections 2.1 and 2.2 it is shown
that the resonance saturation of the sum rule for the amplitude
derivative results in a reasonably small resonance contribution to
the amplitude, whereas the resonance saturation of the sum rule
for the amplitude results in an unacceptably   large resonance
contribution to the amplitude derivative. In Section 2.3 it is
shown that the simultaneous resonance saturation of the amplitude
and its derivative allows to conclude that the resonance
saturation of the sum rule for the amplitude derivative results in
the minimum of $\min\sum_VBR(Z\to\gamma V)$, which agrees
reasonably with the quark model prediction \cite{guberina-80}.
Various deviations from this lower bound are considered.

 In Section 3 there is a brief conclusion.  Specifically, it is
discussed the possibility of $BR(Z\to\gamma J/\psi(1S))\sim
BR(Z\to\gamma\Upsilon(1S))\sim 10^{-6}$, that could be probably
measured in LHC.

In Appendix the angle distributions in the $Z\to\gamma\psi$ and
$Z\to\gamma \Upsilon$ decays are calculated.

\section{Decays  $Z\rightarrow\gamma\psi$ and $Z\rightarrow\gamma\Upsilon$
in dispersion approach}

   As is generally known \cite{rosenberg-63,adler-69} the axial-vector
   vertex,
resulted from the triangle diagrams
\begin{eqnarray}
\label{ACVCVC}
 T_{\alpha \beta \mu} & = &
A_1k_1^{\sigma}\epsilon_{\sigma \alpha \beta \mu}+
A_2k_2^{\sigma}\epsilon_{\sigma \alpha \beta
\mu}+A_3k_{1\beta}k_1^ {\delta}k_2^{\sigma}\epsilon_{\delta \sigma
\alpha \mu}+ \nonumber\\ & & +
A_4k_{2\beta}k_1^{\delta}k_2^{\sigma}\epsilon_{\delta \sigma
\alpha \mu}
+A_5k_{1\alpha}k_1^{\delta}k_2^{\sigma}\epsilon_{\delta \sigma
\beta \mu}+ A_6k_{2\alpha}k_1^{\delta}k_2^{\sigma}\epsilon_{\delta
\sigma \beta \mu}.
\end{eqnarray}
The local gauge invariance
\begin{eqnarray}
k_1^{\alpha}T_{\alpha \beta \mu}=k_2^{\beta}T_{\alpha \beta \mu}=0
\end{eqnarray}
is ensured by the next constraints:
\begin{eqnarray}
A_1=k_2^2A_4+(k_1k_2)A_3,\qquad A_2=k_1^2A_5+(k_1k_2)A_6.
\end{eqnarray}
Besides that
\begin{eqnarray}
A_3(k_1,k_2)=-A_6(k_2,k_1),\qquad A_4(k_1,k_2)=-A_5(k_2,k_1).
\end{eqnarray}
$A_3,A_4,A_5$ and $A_6$ are the invariant amplitudes free of
kinematical singularities. They are well-defined and can be
calculated in the analytic form if $k_1^2=0$ (or $k_2^2=0$). Let
us consider the region $k_1^2=0,\enskip Q^2=-k_2^2=-E^2>0,\enskip
W^2=-M^2=-(k_1+k_2)^2>0$ which is suitable for the calculations
with the help of the dispersion relations over $M^2$ (and over
$E^2$). In the following, it is required only the $A_4$ and $A_6$
amplitudes:
\begin{eqnarray}
\label{BasicFormulae}
 && A_4  =
 -\frac{1}{2\pi^2}\cdot\frac{1}{Q^2-W^2}L_1\,,\nonumber\\
 && A_6=\frac{1}{2\pi^2}\cdot\frac{1}{Q^2-W^2}\left\{\frac{Q^2}{Q^2-
W^2}L_1+\frac{m_q^2}{Q^2-W^2}L_2-1\right\}\,,
\end{eqnarray}
where
\begin{eqnarray}
\label{L1L2}
 L_1 & = & -\rho \ln \frac{\rho+1}{\rho-1} +  \beta
\ln \frac{\beta+1}{\beta-1} \enskip, \nonumber\\ L_2 & = & -\ln^2
\frac{\rho+1}{\rho-1} + \ln^2 \frac{\beta+1}{\beta-1}\enskip,
\nonumber\\ \rho^2 & = & 1+\frac{4m_q^2}{W^2},\qquad
\beta^2=1+\frac{4m_q^2}{Q^2}\enskip.
\end{eqnarray}

   In the other regions of $M^2=-W^2$ and $E^2=-Q^2$ the functions $L_1$ and
$L_2$ are continued analytically in the following way
\cite{achasov-89}\\[6pt]
 i)
\begin{eqnarray}
\label{MContinuation}
 \qquad 0 < -W^2 & = & M^2 < 4m_q^2 \enskip :
\nonumber\\ \rho \rightarrow i\sqrt{-\rho^2} , & & \quad
\frac{1}{2}\ln \frac{\rho+ 1}{\rho-1} \rightarrow -i\arctan
\frac{1}{\sqrt{-\rho^2}}\enskip , \nonumber\\ \qquad 2m_q < M
\enskip & : &  \nonumber\\ \sqrt{-\rho^2} \rightarrow -i\rho , & &
\quad \arctan \frac{1}{\sqrt{- \rho^2}} \rightarrow \frac{\pi}{2}
+ \frac{i}{2}\ln \frac{1+\rho}{1-\rho} .
\end{eqnarray}
ii)
\begin{eqnarray}
\label{EContinuation}
  \quad 0 < -Q^2 & = & E^2 < 4m_q^2
\enskip : \nonumber\\ \beta \rightarrow i\sqrt{-\beta^2} , & &
\quad \frac{1}{2}\ln \frac{\beta+ 1}{\beta-1} \rightarrow
-i\arctan \frac{1}{\sqrt{-\beta^2}}\enskip , \nonumber\\ \quad
2m_q < E \enskip & : & \nonumber\\ \sqrt{-\beta^2} \rightarrow
-i\beta , & & \quad \arctan \frac{1}{\sqrt{- \beta^2}} \rightarrow
\frac{\pi}{2} + \frac{i}{2}\ln \frac{1+\beta}{1-\beta} .
\end{eqnarray}

   It is seen from Eqs.(\ref{BasicFormulae})-(\ref{EContinuation}) that the amplitudes $A_4$ and $A_6$ contain no
singularities (both for $Q^2=-k_2^2\not=0$ and $k_2^2=-Q^2=0$)
except for the dynamical cuts over $4m_q^2\leq M^2<\infty $ and
$4m_q^2\leq E^2<\infty $ resulted from the $q\bar q$ intermediate
states.

   Let us use  Eqs.(\ref{ACVCVC})-(\ref{EContinuation}) to calculate the
amplitude for $Z\rightarrow \gamma(k_1)\gamma^\ast(k_2)$ according
to the  triangle diagrams with the intermediate heavy quarks
$(Z\rightarrow c\bar c\rightarrow \gamma \gamma^\ast$ or
$Z\rightarrow b\bar b\rightarrow \gamma \gamma^\ast)$ for $0 \le
k_2^2 = E^2 \le 4m_q^2\enskip (k_1^2 = 0)$ in the rest frame of
the $Z$ boson.
\begin{eqnarray}
\label{Ztogg*1}
 && T\Bigl (Z\to q\bar q\to\gamma \gamma^\ast\Bigr )
= M^2E\left(1 - E^2/M^2\right)t_q\times\nonumber\\
 &&\times\biggl\{\Bigl (\frac{E}{M}\Bigr )\Bigl (\mbox{\bf n}\cdot\mbox{\bf
e($Z$)}\Bigr )\Bigl (\mbox{\bf n}\cdot\Bigl [\mbox{\bf
e($\gamma^\ast$)}\times\mbox{\bf e($\gamma$)}\Bigr ]\Bigr ) +
\Bigl (\mbox{\bf n}\cdot\mbox{\bf e($\gamma^\ast$)}\Bigr )\Bigl
(\mbox{\bf n}\cdot\Bigl [\mbox{\bf e($\gamma$)}\times\mbox{\bf
e($Z$)}\Bigr ]\Bigl)\biggr\}\,,
\end{eqnarray}
where $M\equiv M_Z$ is the mass of the $Z$ boson, $M^2 = (k_1 +
k_2)^2$ ; $\mbox{\bf n} = \mbox{\bf k$_1$}/\Bigl |\mbox{\bf
k$_1$}\Bigr |$ ; {\bf e($Z$)} and {\bf e($\gamma^\ast$)} are the
polarization three-vectors of the $Z$ boson and the $\gamma^\ast$
quantum in their rest frames ; {\bf e($\gamma$)} is the
polarization three-vector of the $\gamma $ quantum. The amplitude
$t_q$ takes into account three identical loops corresponding to
three colors.
\begin{equation}
\label{tq}
 t_q = -\sigma_q \frac{3}{4}\cdot \frac{e^3e_q^2}{\sin
2\Theta_W}\left(A_4 + A_6\right) ,
\end{equation}
where $\sigma_c = 1, \sigma_b = -1, e_c = 2/3, e_b = -1/3.$

   It is seen from Eqs.(\ref{BasicFormulae})-(\ref{EContinuation}) that $t_q$ satisfies a  dispersion
relation without subtractions  both in  $M^2$ and in $E^2$.
Consequently, $t_q$ is the amplitude convenient for obtaining sum
rules in the $E^2$ channel. Since at the present time it appears
to test  theoretically only the resonance saturation of the sum
rules evaluated below, it is most convenient to derive them with
the help of the following consideration.

    The amplitude $t_q$ describes the full  amplitude for $Z\rightarrow
q\bar q\rightarrow \gamma \gamma^\ast$ in the region $E^2\le 0$
accurate up to
 higher corrections in $QCD$ and the standard
electroweak theory, i.e., accurate up to  corrections of  order
$\alpha_S(4m_q^2)/\pi,\enskip \alpha_S(M^2)/\pi$ and $\alpha/\pi$.
On the other hand, the full  amplitude for $Z\rightarrow q\bar
q\rightarrow \gamma \gamma^\ast$ can be represented with the help
of the intermediate hadronic  states in the $E^2$ channel as the
sum of  resonance  contributions and a continuum spectrum
contribution:
\begin{eqnarray}
\label{Ztogg*2}
 && T\Bigl (Z\to q\bar q\rightarrow \gamma \gamma^\ast\Bigr )
= M^2E\left(1 - E^2/M^2\right)t^q_h\times\nonumber\\
&&\times\biggl\{\Bigl (\frac{E}{M}\Bigr )\Bigl (\mbox{\bf
n}\cdot\mbox{\bf e($Z$)}\Bigr )\Bigl (\mbox{\bf n}\cdot\Bigl
[\mbox{\bf e($\gamma^\ast$)}\times\mbox{\bf e($\gamma$)}\Bigr
]\Bigr ) + \Bigl (\mbox{\bf n}\cdot\mbox{\bf
e($\gamma^\ast$)}\Bigr )\Bigl (\mbox{\bf n}\cdot\Bigl [\mbox{\bf
e($\gamma$)}\times\mbox{\bf e($Z$)}\Bigr ]\Bigl)\biggr\}\,,
\end{eqnarray}
where
\begin{equation}
\label{thq}
 t_h^q = \sum_{V}\frac{m_V^2}{m_V^2 - E^2}\cdot
\frac{e}{f_V}T_V^q + eT_{cont}^q.
\end{equation}
\begin{sloppypar}
Here $V$ is a $(q\bar q)$ vector quarkonium ; $T_{cont}^q$ is the continuum
contribution ($D\bar D, D^\ast\bar D, D\bar D^\ast, D^\ast\bar D^\ast,\cdots$
or $B\bar B, B^\ast\bar B, B\bar B^\ast, B^\ast\bar B^\ast,\cdots$).
\end{sloppypar}
   There is every reason to believe that where $E^2\approx 0$
\begin{equation}
\label{qhduality}
 t_h^q \approx t_q \approx -\sigma_q\frac{3e\alpha e_q^2}{2\sin
2\Theta_W}\cdot \frac{1}{M^2}\left(i - \frac{2}{\pi}\ln
\frac{M}{m_q} + \frac{1}{\pi} + \frac{\beta}{\pi}\ln \frac{\beta +
1}{\beta - 1}\right).
\end{equation}
 Eq.(\ref{qhduality}) incorporates the fact  that $2m_q/M\ll 1$.

   Let us consider at $E^2=0$ the sum rule for the amplitude
\begin{equation}
\label{sra}
 t_q\biggl|_{E^2=0} = t_h^q\biggr|_{E^2=0}
\end{equation}
and its first derivative
\begin{equation}
\label{srda}
\frac{d}{dE^2}t_q\Biggl|_{E^2=0}=
\frac{d}{dE^2}t_h^q\Biggr|_{E^2=0}.
\end{equation}

   It follows from Eqs.(\ref{thq}), (\ref{qhduality}) and (\ref{sra}) that
\begin{eqnarray}
\label{sra1}
 & & \sum_{V}\frac{1}{f_V}T_V^q +
T_{cont}^q\biggl|_{E^2=0}\equiv T_q(Res) +
T_{cont}^q\biggr|_{E^2=0}= \nonumber\\
 & & = T_q \equiv
-\sigma_q\frac{3\alpha e_q^2}{2\sin 2\Theta_W}\cdot \frac{1}{M^2}
\left(i - \frac{2}{\pi}\ln \frac{M}{m_q} + \frac{3}{\pi}\right).
\end{eqnarray}

   An unusual feature of this sum rule is the presence of an imaginary part on
the right-hand side of  Eq. (\ref{sra1}) coming from the amplitude
discontinuity due to the real $q\bar q$ intermediate states in the
$M^2$ channel \footnote{The contribution of resonances in
$Im(T_q)$ should be small for the vertex $ q
^*(\mbox{virtual})\bar q\to V $ (or $\bar q^*(\mbox {virtual})
q\to V $) should be suppressed by the wave function of the
quarkonium. This reason was missed in Refs.
\cite{achasov-Z91,achasov-Z92}.}.

   It follows from Eqs. (\ref{thq}), (\ref{qhduality}) and (\ref{srda}) that
\begin{eqnarray}
\label{srda1}
 & & \sum_{V}\frac{1}{f_Vm_V^2}T_V^q +
\frac{d}{dE^2}T_{cont}^q\Biggl|_{E^2=0} = D_q(Res) +
\frac{d}{dE^2}T_{cont}^q\Biggr|_{E^2=0} = \nonumber\\ & & = D_q
\equiv  \sigma_q \frac{\alpha e_q^2}{4\pi m_q^2\sin
2\Theta_W}\cdot \frac{1}{M^2}.
\end{eqnarray}

$ImD_q = 0$ for the approximation (\ref{qhduality}).

   The width of the decay $Z\rightarrow \gamma V$ is
\begin{eqnarray}
\label{GZtogV}
 & & \Gamma (Z\rightarrow \gamma V) = \nonumber\\
 & & =
\frac{1}{24\pi}(1 - m_V^2/M^2)^3(1 +
m_V^2/M^2)M^3m_V^2\left|T_V^q\right|^2\approx\frac{1}{24\pi}
M^3m_V^2\left|T_V^q\right|^2.
\end{eqnarray}

   To determine $f_V^2/4\pi$ , one uses the experimental data \cite{pdg-10} on
\begin{equation}
\label{GVtoe+e-}
 \Gamma (V\rightarrow e^+e^-) =
\frac{4\pi}{3}\frac{m_V}{f_V^2}\alpha^2
\end{equation}
As a result one gets for the $\psi$ family $c1\equiv
J/\psi(1S)\equiv\psi (3097),\ c2\equiv \psi (3686),\ c3\equiv \psi
(3770),\ c4\equiv \psi (4040),\ c5\equiv \psi (4160),\ c6\equiv
\psi (4415)$,
\begin{equation}
\label{fc}
 f_{c1}:f_{c2}:f_{c3}:f_{c4}:f_{c5}:f_{c6} = 1:1.67:5.05:2.9:3:3.69\,,
\end{equation}
and for the $\Upsilon$ family $b1\equiv \Upsilon (9460),\ b2\equiv
\Upsilon (10023),\ b3\equiv \Upsilon (10355),\ b4\equiv \Upsilon
(10579),\ b5\equiv \Upsilon (10860),\ b6\equiv \Upsilon (11020)$,
\begin{equation}
\label{fb}
 f_{b1}:f_{b2}:f_{b3}:f_{b4}: f_{b5}:f_{b6} =
1:1.52:1.82:2.35:2.23:3.47\,,
\end{equation}
where
\begin{eqnarray}
\label{groundfs} && f_{c1}=11.16,\ \ \ f_{c1}^2/4\pi\equiv
f_{J/\psi(1S)}^2/4\pi\equiv f_{\psi(3097)}^2/4\pi =9.91\,,
\nonumber\\[6pt]
 && f_{b1}= 39.69\,,\ \ \ f_{b1}^2/4\pi\equiv
f_{\Upsilon(1S)}^2/4\pi\equiv  f_{\Upsilon(9460)}^2/4\pi
 = 125.38\,.
\end{eqnarray}

\subsection{Sum rule for amplitude}

   Let us assume initially that the real part
  of the sum rule for the
amplitude, Eq. (\ref{sra1}), is saturated with a ground state,
that is,
\begin{equation}
\label{sragrounds}
  T_V^q = f_VRe(T_q)\,,\ \ \mbox{where}\,,\
V = J/\psi(1S),\enskip \Upsilon (1S).
\end{equation}

   Using Eqs. (\ref{sra1}), (\ref{GZtogV}), (\ref{groundfs}), (\ref{sragrounds}), $m_c=1.27$ GeV, $m_b=4.2$ GeV, $M=91.19$ GeV,
$\Gamma_Z=2.5$ GeV, $\alpha=1/137$, and $\sin 2\Theta_W=0.84$,
\cite{pdg-10}, we find
\begin{equation}
\label{BZtogJPsiU1Ssra}
 BR(Z\rightarrow \gamma J/\psi(1S)) = 7.2\cdot 10^{-6}, \quad
BR(Z\rightarrow \gamma \Upsilon (1S)) = 1.7\cdot 10^{-5},
\end{equation}
which are two orders of magnitude higher than it is expected  in
the quark model \cite{guberina-80}.

  Let us saturate now  the real part
   of the sum rule for the
amplitude Eq. (\ref{sra1}) with the $\psi$ and $\Upsilon$ families
\begin{equation}
\label{sraresonances}
 \sum_V\frac{1}{f_V}T_V^q \equiv T_q(Res) = Re(T_q)\,.
\end{equation}
Note that this takes partially into account  the continuous
spectrum since  four members of the $\psi$ family and three
members of the $\Upsilon$ family lie  in the continuous spectrum
of $D\bar D, D\bar D^\ast, D^\ast\bar D, \bar D^\ast D^\ast$ and
$B\bar B, B\bar B^\ast, B^\ast\bar B, \bar B^\ast B^\ast$,
respectively.

   Considering Eq. (\ref{sraresonances}) as the constraint and using Eq. (\ref{GZtogV})
   one can find $\min \sum_V\Gamma (Z\rightarrow \gamma V)$ that is reached when
   \begin{equation}
\label{individualminsraresonanceamplitude}
  T_V^q =\frac{1}{a_qf_Vm_V^2}Re(T_q)\,,\ \ \ \mbox{where}\ \ \
 a_q  =  \sum_V\frac{1}{f_V^2m_V^2}\,,
\end{equation}
\begin{equation}
\label{individualminsraresonances}
 \Gamma (Z\rightarrow \gamma V) = \frac{1}{24\pi}M^3\left (Re(T_q)\right )^2\left(f_V m_V
 a_q\right)^{-2}\,,
\end{equation}
and
\begin{eqnarray}
\label{minsraresonances}
 && \min \sum_V\Gamma (Z\rightarrow \gamma
V)  =  \frac{1}{24\pi}M^3\left (Re(T_q)\right )^2a_q^{-1}\,.
\end{eqnarray}

For the $\psi$ family ($a_c = 1.2\cdot 10^{-3}\ \mbox{GeV}^{-2}$)
\begin{eqnarray}
\label{cJpsiminsraresonances} && \min \sum_{\Psi} BR(Z\rightarrow
\gamma \psi)  =  5.05\cdot 10^{-6},
 \nonumber\\[6pt] &&
BR(Z\rightarrow \gamma J/\psi(1S))  =  3.53\cdot 10^{-6}.
\end{eqnarray}
For the $\Upsilon$ family ($a_b = 1.43\cdot 10^{-5}\
\mbox{GeV}^{-2}$)
\begin{eqnarray}
\label{bU1Siminsraresonances}
 && \min \sum_{\Upsilon}
BR(Z\rightarrow \gamma \Upsilon)  =  8.58\cdot 10^{-6},
\nonumber\\[6pt]
&& BR(Z\rightarrow \gamma \Upsilon (1S)  =
4.25\cdot 10^{-6}.
\end{eqnarray}

When the amplitude is saturated with the ground state or the
resonance family, it follows from Eqs. (\ref{sragrounds}) or
(\ref{sraresonances})
\begin{equation}
\label{DqVTq}
 D_q(V)=\frac{1}{f_Vm_V^2}T_V^q=\frac{1}{m^2_V}Re(T_q)
\end{equation}
or
\begin{equation}
\label{DqTq}
 D_q(Res)=\sum_V\frac{1}{f_Vm_V^2}T_V^q=\sum_V\frac{1}{f_Vm_V^2}\cdot\frac{1}{a_qf_Vm^2_V}Re(
 T_q)=\frac{d_q}{a_q}Re(T_q)\,.
\end{equation}

Eq. (\ref{DqVTq}) leads to
\begin{equation}
 \label{DcJPsiTc}
 D_c(J/\psi(1S))=0.10Re(T_c)\ \mbox{GeV}^{-2}=5.60D_c
\end{equation}
and
\begin{equation}
 \label{DbUpsilon1STb}
 D_b(\Upsilon (1S))=0.01Re(T_b)\ \mbox{GeV}^{-2}= 3.73D_b\,.
\end{equation}

Eq. (\ref{DqTq}) leads to
\begin{equation}
\label{DcTc}
 D_c(Res)=0.09Re(T_c)\ \mbox{GeV}^{-2}=4.99D_c
\end{equation}
and
\begin{equation}
\label{DbTb}
 D_b(Res)=0.01Re(T_b)\ \mbox{GeV}^{-2}=3.42D_b\,.
\end{equation}

So, the saturation of the amplitudes with the ground states or the
resonance families leads to the unacceptably large contributions
of the resonances into the amplitude derivatives\footnote{This
crucial point was missed in Refs.
\cite{achasov-Z91,achasov-Z92}.}.

   Using Eqs. (\ref{BasicFormulae})-(\ref{EContinuation}) one can verify that the dispersion integral
for $T_q$ is determined by the region $2m_q\le E\sim M_Z$, which
is hardly  a low energy region. Consequently, it is reasonable to
study the sum rule (\ref{srda1}) for the amplitude derivative
because the contribution of low-lying states in the dispersion
integral for the amplitude derivative is significantly enhanced as
compared to their contribution to the amplitude itself.  Note that
90\% of the dispersion integral for $D_q$  is determined by the
region of low energies $2m_q\le E \le 6m_q$.

\subsection{Sum rule for the  derivative of  the amplitude}
 Let us assume initially that  the sum rule for the amplitude derivative, Eq. (\ref{srda1}),
 is saturated with a ground state, $V = J/\psi(1S),\enskip \Upsilon(1S)$,
  \begin{equation}
\label{srdagrounds}
  T_V^q = f_Vm^2_VD_q\,,
 \end{equation}
 then
 \begin{equation}
  \Gamma (Z\to\gamma V) =\frac{1}{24\pi}
f^2_VM^3m_V^6D_q^2\,,
\end{equation}
resulting in
\begin{equation}
\label{BZtogJPsiU1Ssrda}
 BR(Z\rightarrow \gamma J/\psi(1S)) = 2.31\cdot 10^{-7}, \quad
BR(Z\rightarrow \gamma \Upsilon (1S)) = 1.24\cdot 10^{-6}.
\end{equation}

As this takes place,  the ground state contribution in $T_q$ is
\begin{equation}
\label{TqVDq}
 T_q(V)\equiv\frac{1}{f_V}T_V^q = m^2_VD_q\,,
\end{equation}
resulting in
\begin{equation}
\label{TcJPsiDc} T_c(J/\psi(1S))=9.59D_c\ \mbox{GeV}^2=0.18Re(T_c)
\end{equation}
and
\begin{equation}
\label{TbUpsilon1SDb} T_b(\Upsilon(1S))=89.49D_b\
\mbox{GeV}^2=0.27Re(T_b)\,.
\end{equation}

 Now let us consider the saturation of  of the sum
rule for the amplitude derivative, Eq. (\ref{srda1}), with the
$\psi$ and $\Upsilon$ families
\begin{equation}
\label{srdaresonances}
 \sum_V\frac{1}{f_Vm_V^2}T_V^q \equiv D_q(Res) = D_q
\end{equation}

  Considering Eq. (\ref{srdaresonances}) as the constraint and using Eq. (\ref{GZtogV})
   one get that $\min \sum_V\Gamma (Z\rightarrow \gamma V)$  is reached when
   \begin{equation}
\label{individualminsrdaresonanceamplitude}
  T_V^q =\frac{1}{g_qf_Vm_V^4}D_q\,,\ \ \ \mbox{where}\ \ \
 g_q  =  \sum_V\frac{1}{f_V^2m_V^6}\,,
\end{equation}
\begin{equation}
\label{individualminsrdaresonances}
 \Gamma (Z\rightarrow \gamma V) = \frac{1}{24\pi}M^3D_q^2\left(f_V m_V^3
 g_q\right)^{-2}\,,
\end{equation}
and
\begin{equation}
\label{minsrdaresonances}
  \min \sum_V\Gamma (Z\rightarrow \gamma
V)  =  \frac{1}{24\pi}M^3D_q^2 g_q^{-1}\,.
\end{equation}

For the $\psi$ family ($g_c =1.08\cdot 10^{-5}\ \mbox{GeV}^{-6}$)
\begin{equation}
\label{cminsrdaresonances}
 \min \sum_{\psi} BR(Z\rightarrow
\gamma \psi)  =  1.95\cdot 10^{-7}
\end{equation}
and
\begin{eqnarray}
\label{psissrdamin}
 && BR(Z\rightarrow \gamma J/\psi(1S) )  =  1.64\cdot
10^{-7}\,,\ \  BR(Z\rightarrow \gamma \psi(3686))  = 2.056\cdot
10^{-8}\,,\nonumber\\[6pt]
 &&
BR(Z\rightarrow \gamma \psi(3770))  =  2\cdot 10^{-9}\,,\ \
BR(Z\rightarrow \gamma \psi(4040))  =  4\cdot
10^{-9}\,,\nonumber\\[6pt]
 &&
BR(Z\rightarrow \gamma \psi(4160))  =  3\cdot 10^{-9}\,,\ \
BR(Z\rightarrow \gamma \psi(4415))  =  1.44\cdot 10^{-9}\,.
\end{eqnarray}
 For the $\Upsilon$ family ($g_b = 1.52\cdot 10^{-9}\
\mbox{GeV}^{-6}$)
\begin{equation}
\label{bminsrdaresonances}
 \min \sum_{\Upsilon}
BR(Z\rightarrow \gamma \Upsilon)  =  7.23\cdot 10^{-7}
\end{equation}
and
\begin{eqnarray}
\label{upsilonssrdamin}
 && BR(Z\rightarrow \gamma
\Upsilon (1S)) = 4.27\cdot 10^{-7}\,,\ \
 BR(Z\rightarrow \gamma \Upsilon (10023)) = 1.31\cdot
10^{-7}\,,\nonumber\\[6pt]
 && BR(Z\rightarrow \gamma \Upsilon (10355)) = 7.5\cdot
10^{-8}\,,\ \ BR(Z\rightarrow \gamma \Upsilon (10579)) = 3.9\cdot
10^{-8}\,,\nonumber\\[6pt]
 &&
BR(Z\rightarrow \gamma \Upsilon (10860)) = 3.7\cdot 10^{-8}\,,\ \
BR(Z\rightarrow \gamma \Upsilon (11020)) = 1.4\cdot 10^{-8}.
\end{eqnarray}

The branching ratios for the production of the ground states
 ( $BR(Z\rightarrow \gamma J/\psi(1S) )$ and $BR(Z\rightarrow \gamma
\Upsilon (1S))$ in Eqs. (\ref{psissrdamin}) and
(\ref{upsilonssrdamin}) )  more or less agree with the quark model
predictions, $\sim 3.4\cdot (10^{-8}-10^{-7})$,
\cite{guberina-80}.

 It follows from Eq.
(\ref{individualminsrdaresonanceamplitude}) that
\begin{equation}
\label{TqDq}
 T_q(Res)\equiv \sum_V\frac{1}{f_V}T_V^q =
\frac{d_q}{g_q}D_q(Res)=\frac{d_q}{g_q}D_q\,,\ \ \mbox{where}\ \
d_q  =  \sum_V\frac{1}{f_V^2m_V^4}\,.
\end{equation}
For the $\psi$ family ($d_c =1.12\cdot 10^{-4}\ \mbox{GeV}^{-4}$)
\begin{equation}
\label{TcDc}
 T_c(Res)\equiv \sum_V\frac{1}{f_V}T_V^c = 10.19
D_c(Res)\ \mbox{GeV}^2=10.19 D_c\ \mbox{GeV}^2=0.19Re(T_c)\,.
\end{equation}
For the $\Upsilon$ family ($d_b =1.47\cdot 10^{-7}\
\mbox{GeV}^{-4}$)
\begin{equation}
\label{TbDb}
 T_b(Res)\equiv \sum_V\frac{1}{f_V}T_V^b = 96.71
D_b(Res)\ \mbox{GeV}^2 = 96.71 D_b\ \mbox{GeV}^2=0.29Re(T_b)\,.
\end{equation}

Eqs. (\ref{TcJPsiDc}), (\ref{TbUpsilon1SDb}) and (\ref{TcDc}),
(\ref{TbDb}) specify explicitly that the main body of $T_c$ and
$T_b$ is saturated with the continuous spectrum, see Eq.
(\ref{sra1}).\footnote{This point was considered in Refs.
\cite{achasov-Z91,achasov-Z92} only partly.}
 In
addition, Eqs. (\ref{TcJPsiDc}), (\ref{TbUpsilon1SDb}) and
(\ref{TcDc}), (\ref{TbDb}) corroborate the  comment in the
footnote 3.

 \subsection{ Sum rules for the amplitude and its derivative}
 The simultaneous saturation of the
amplitude and its derivative with the ground state is provided  if
only
\begin{equation}
\label{srasrdaground}
 Re(T_q)/D_q = m^2_V\,,
\end{equation}
but in our case
\begin{eqnarray}
\label{srasrdaJPsiUpsilon1S} && Re(T_c)/D_c=53.688\
\mbox{GeV}^2\neq  m^2_{J/\psi(1S)}=9.59\ \mbox{GeV}^2\ \ \
\mbox{and}\nonumber\\ && Re(T_b)/D_b=334\ \mbox{GeV}^2\neq
 m^2_{\Upsilon (1S)}=89.49\ \mbox{GeV}^2\,.
\end{eqnarray}

As for the simultaneous saturation of the amplitude and its
derivative with the resonance family, it's quite another matter.

   Considering the resonance contributions  in the sum rules for the
amplitude, $T_q(Res)$, and its derivative, $D_q(Res)$,  as the two
constraints and using Eq. (\ref{GZtogV}) we find
\begin{eqnarray}
\label{twoconstraintsmin}
 && \min \sum_V \Gamma (Z\rightarrow \gamma
V) = \nonumber\\[6pt]
 &&=
\frac{1}{24\pi}M^3\cdot\frac{g_qT_q(Res^2)^2 +
 a_qD_q(Res)^2 - 2d_qT_q(Res)D_q(Res)}{a_q g_q - d_q^2}\,.
\end{eqnarray}

Eq. (\ref{twoconstraintsmin}) takes place when
\begin{equation}
\label{twoconstraintsminamp} T_V^q = \frac{1}{f_V m_V^2}\cdot
\frac{(g_q - d_q/m_V^2)T_q(Res) - (d_q - a_q/m_V^2)D_q(Res)}{a_q
g_q - d_q^2}.
\end{equation}
It is easy to verify that Eq. (\ref{twoconstraintsminamp}) is
self-consistent for any $T_q(Res)$,  $D_q(Res)$, and $m_V^2$.
\begin{equation}
\label{verify}
 \sum_V\frac{1}{f_V}T_V^q=T_q(Res)\,,\ \ \
 \sum_V\frac{1}{f_Vm^2_V}T_V^q=D_q(Res)\,.
\end{equation}

 The minimum of Eq. (\ref{twoconstraintsmin}) takes place when
\begin{equation}
\label{TqDqRe}
 T_q(Res) = \frac{d_q}{g_q}D_q(Res).
\end{equation}

Setting $D_q(Res)=D_q$, we revert to the previous subsection, to
the saturation of the amplitude derivative with  the resonance
family.

Let us consider the deviation from Eq. (\ref{TqDqRe})\
\footnote{The deviations from the  lower bounds of $\sum_{\psi}
BR(Z\to\gamma\psi)$ and $\sum_{\Upsilon} BR(Z\to\gamma\Upsilon)$
were not studied  in Refs. \cite{achasov-Z91,achasov-Z92} in the
regular way.}
\begin{equation}
\label{TqDqDeviation}
 T_q(Res) = \frac{d_q}{g_q}D_q(Res)\cdot (1+x)=\frac{d_q}{g_q}D_q\cdot (1+x)\,,
\end{equation}
then
\begin{equation}
\label{twoconstraintsminresonances}
  \min \sum_V\Gamma (Z\rightarrow \gamma V)=
 \frac{1}{24\pi}M^3D_q^2
g_q^{-1}\left (1+x^2\cdot\frac{d_q^2}{\Delta_q}\right )\,,
 \end{equation}
 where $\Delta_q=a_qg_q-d_q^2$,
   \begin{equation}
\label{individualamptwoconstraintsmin}
  T_V^q =\frac{1}{g_qf_Vm_V^4}D_q\left [1+x\cdot\frac{d_q\left (g_qm_V^2-d_q\right )}{\Delta_q}\right ]\,,
\end{equation}
and
\begin{eqnarray}
\label{individualwidthtwoconstraintsmin} && \Gamma (Z\rightarrow
\gamma V) =\nonumber\\[6pt]
 && = \frac{1}{24\pi}M^3D_q^2\left(f_V
m_V^3 g_q\right)^{-2}
 \left [1+ 2x\cdot\frac{d_q\left (g_qm_V^2-d_q\right )}{\Delta_q}+
  x^2\cdot\frac{d_q^2\left (g_qm_V^2-d_q\right )^2}{\Delta_q^2}\right
  ]\,.
  \end{eqnarray}
It is easy to verify that the term, proportional $x$ in Eq.
(\ref{individualwidthtwoconstraintsmin}), vanishes in
$$\sum_V\Gamma (Z\rightarrow \gamma V)\,.$$
 For the $\psi$ family ($\Delta_c = 4.45\cdot 10^{-10}\ \mbox{GeV}^{-8}$)
 \begin{equation}
 \label{psifamilytwoconstraintsmin}
  \min \sum_{\psi} BR(Z\rightarrow
\gamma \psi)  =  1.95\cdot 10^{-7}\cdot (1+x^2\cdot 28.17 )
 \end{equation}
 and
\begin{eqnarray}
\label{psistwoconstraintsmin}
 && BR(Z\rightarrow \gamma J/\psi(1S) )  =  1.64\cdot
10^{-7}\cdot (1-x\cdot 4.29+x^2\cdot 4.60)\,,\nonumber\\[6pt]
 &&
BR(Z\rightarrow \gamma\psi(3686))  =  2.056\cdot 10^{-8}\cdot
(1+x\cdot 17.392+x^2\cdot 75.62)\,,\nonumber\\[6pt]
 &&
BR(Z\rightarrow \gamma\psi(3770))  =  2\cdot 10^{-9}\cdot
(1+x\cdot 20.79+x^2\cdot 108.07)\,,\nonumber\\[6pt]
 &&
BR(Z\rightarrow \gamma\psi(4040))  =  4\cdot 10^{-9}\cdot
(1+x\cdot 32.236+x^2\cdot 259.79)\,,\nonumber\\[6pt]
 &&
BR(Z\rightarrow \gamma\psi(4160))  =  3\cdot 10^{-9}\cdot
(1+x\cdot 37.58+x^2\cdot 353)\,,\nonumber\\[6pt]
 &&
BR(Z\rightarrow \gamma\psi(4415))  =  1.44\cdot 10^{-9}\cdot
(1+x\cdot 49.44+x^2\cdot 611.19)\,.
\end{eqnarray}
For the $\Upsilon$ family ($\Delta_b = 2.14\cdot 10^{-16}\
\mbox{GeV}^{-8}$)
\begin{equation}
\label{upsilonfamilytwoconstraintsmin}
  \min \sum_{\Upsilon} BR(Z\rightarrow \gamma
\Upsilon)  =  7.23\cdot 10^{-7}\cdot (1+x^2\cdot 100.46)
\end{equation}
and
\begin{eqnarray}
\label{upsilonstwoconstraintsmin}
 && BR(Z\rightarrow \gamma
\Upsilon (1S)) = 4.27\cdot 10^{-7}\cdot (1- x\cdot 15.04 +x^2\cdot
56.57)\,,\nonumber\\[6pt]
 && BR(Z\rightarrow \gamma \Upsilon (10023)) = 1.31\cdot
10^{-7}\cdot (1+ x\cdot 7.74 +x^2\cdot 14.97 )\,,\nonumber\\[6pt]
 && BR(Z\rightarrow \gamma \Upsilon (10355)) = 7.5\cdot
10^{-8}\cdot (1+ x\cdot 21.79 +x^2\cdot 118.71)\,,\nonumber\\[6pt]
&& BR(Z\rightarrow \gamma \Upsilon (10579)) = 3.9\cdot
10^{-8}\cdot (1+ x\cdot 31.53 +x^2\cdot 248.54)\,,\nonumber\\[6pt]
 &&
BR(Z\rightarrow \gamma \Upsilon (10860)) = 3.7\cdot 10^{-8}\cdot
(1+ x\cdot 44.04 +x^2\cdot 484.88)\,,\nonumber\\[6pt]
 &&
BR(Z\rightarrow \gamma \Upsilon (11020)) = 1.4\cdot 10^{-8}\cdot
(1+ x\cdot 51.31 +x^2\cdot 658.27).
\end{eqnarray}

 When the resonances saturate $T_q$,  $x=4.26$ for the $\psi$ family
 and $x=2.45$ for the $\Upsilon$ one. As this takes place,
 \begin{eqnarray}
 \label{saturationDcTc}
 && \sum_\psi BR(Z\to\gamma\psi)=10^{-4}\,,\nonumber\\
 &&  BR(Z\to\gamma J/\psi(1S))=1.1\cdot 10^{-5}\,,\ \ BR(Z\to\gamma\psi(3686))=3\cdot
 10^{-5}\,,\nonumber\\
 && BR(Z\to\gamma\psi(3770))=4\cdot 10^{-6}\,,\ \ BR(Z\to\gamma\psi(4040))=1.9\cdot
  10^{-5}\,,\nonumber\\
  &&  BR(Z\to\gamma\psi(4160))=2\cdot 10^{-5}\,,\ \  BR(Z\to\gamma\psi(4415))=1.6\cdot 10^{-5}
 \end{eqnarray}
and
\begin{eqnarray}
 \label{saturationDbTb}
 && \sum_\Upsilon BR(Z\to\gamma\Upsilon)=4.36\cdot 10^{-4}\,,\nonumber\\
 &&  BR(Z\to\gamma
 \Upsilon(1S))=1.28\cdot 10^{-4}\,,\ \ BR(Z\to\gamma\Upsilon(10023))=1.5\cdot
 10^{-5}\,,\nonumber\\
 && BR(Z\to\gamma\Upsilon(10355))=5.9\cdot 10^{-5}\,,\ \ BR(Z\to\gamma\Upsilon(10579))=6.3\cdot
  10^{-5}\,,\nonumber\\
  &&  BR(Z\to\gamma\Upsilon(10860))=1.12\cdot 10^{-4}\,,\ \  BR(Z\to\gamma\Upsilon(11020))=5.9\cdot
  10^{-5}\,.
\end{eqnarray}

As noted in Section 2.1, there are no theoretical grounds for
saturating $T_q$ with the resonances only. Furthermore the
prediction  $BR(Z\to\gamma\Upsilon(1S))=1.28\cdot 10^{-4}$, see
Eq. (\ref{saturationDbTb}), contradicts the experiment value
$BR(Z\to\Upsilon(1S)X)<4.4\cdot 10^{-5}$ on CL=95 \% \cite{pdg-10}
thus bearing the theoretical reason.

When $x=-1$, the resonances do not contribute to $T_q$ at all. As
this takes place,
 \begin{eqnarray}
 \label{saturationDconly}
 && \sum_\psi BR(Z\to\gamma\psi)= 5.69\cdot 10^{-6}\,,\nonumber\\
 &&  BR(Z\to\gamma J/\psi(1S))=1.62\cdot 10^{-6}\,,\ \ BR(Z\to\gamma\psi(3686))=1.22\cdot
 10^{-6}\,,\nonumber\\
 && BR(Z\to\gamma\psi(3770))=1.7\cdot 10^{-7}\,,\ \ BR(Z\to\gamma\psi(4040))=9\cdot
  10^{-7}\,,\nonumber\\
  &&  BR(Z\to\gamma\psi(4160))=9.8\cdot 10^{-7}\,,\ \  BR(Z\to\gamma\psi(4415))=8\cdot 10^{-7}
 \end{eqnarray}
 and
\begin{eqnarray}
 \label{saturationDbonly}
 && \sum_\Upsilon BR(Z\to\gamma\Upsilon)=7.34\cdot 10^{-5}\,,\nonumber\\
 &&  BR(Z\to\gamma
 \Upsilon(1S))=3.08\cdot 10^{-5}\,,\ \ BR(Z\to\gamma\Upsilon(10023))=1.3\cdot
 10^{-6}\,,\nonumber\\
 && BR(Z\to\gamma\Upsilon(10355))=7.4\cdot 10^{-6}\,,\ \ BR(Z\to\gamma\Upsilon(10579))=8.7\cdot
  10^{-6}\,,\nonumber\\
  &&  BR(Z\to\gamma\Upsilon(10860))=1.65\cdot 10^{-5}\,,\ \  BR(Z\to\gamma\Upsilon(11020))=8.7\cdot
  10^{-6}\,.
\end{eqnarray}

Zeros in $BR(Z\to\gamma J/\psi(1S)$ at $x=0.466$ and in
$BR(Z\to\gamma\Upsilon(1S)$ at $x=0.133$ are striking, see Eqs.
(\ref{psistwoconstraintsmin}) and
(\ref{upsilonstwoconstraintsmin}). In  this case
\begin{equation}
\label{zeros}
 \sum_{\psi\neq J/\psi} BR(Z\to\gamma\psi)=1.39\cdot
10^{-6}\,,\ \ \sum_{\Upsilon\neq \Upsilon (1S)}
BR(Z\to\gamma\Upsilon)=2.01\cdot 10^{-6}\,,
\end{equation}
and
\begin{equation}
\label{zerosSaturation}
 T_c(Res)=0.28T_c\,,\ \ T_b(Res)=0.33T_b\,.
\end{equation}

It follows from Eq. (\ref{zerosSaturation}) that the continues
spectra dominate the saturation of the $T_c$ and $T_b$ amplitudes,
but zeros, see Eq. (\ref{individualamptwoconstraintsmin}),
\begin{equation}
\label{zerosTqV} T^c_{J/\psi(1S)}|_{x=0.466}=0\,\ \mbox{and}\ \
T^b_{\Upsilon(1S)}|_{x=0.133}=0
\end{equation}
require a rather  bizarre dynamics, as I believe.

\section{Summary}
As is evident from the foregoing, the lower bounds of $\sum_{\psi}
BR(Z\to\gamma\psi) =  1.95\cdot 10^{-7}$ and $\sum_{\Upsilon}
BR(Z\to\gamma\Upsilon) =  7.23\cdot 10^{-7}$ are reached when the
 derivatives of the
$Z\to c\bar c\to\gamma\gamma^{\ast}$ and $Z\to b\bar b\to\gamma
\gamma^{\ast}$ amplitudes are saturated with the resonances in the
 $\gamma^{\ast}$ low energy region. As this takes place, the
branching ratios for the production of the ground states,
 $BR(Z\to\gamma J/\psi(1S) )=1.64\cdot
10^{-7}$ and $BR(Z\to\gamma\Upsilon (1S))=4.27\cdot 10^{-7}$, more
or less agree with the quark model predictions.

These lower bounds  are  "the equilibrium points " of Eqs.
(\ref{psifamilytwoconstraintsmin}) and
(\ref{upsilonfamilytwoconstraintsmin}) at $x=0$.  The minima are
rather sharp, especially in the $\Upsilon$ family case. Thus for
the 20 percentage  reduction of the   $\Upsilon$ family
contribution to the triangle diagram amplitude,  $x= -0.2$, $
\Sigma_\Upsilon BR(Z\to\gamma \Upsilon) = 3.6\times 10^{-6}$ and
$BR(Z\to\gamma \Upsilon(1S)= 2.7\times 10^{-6}$ are resulted from
Eqs. (\ref{upsilonfamilytwoconstraintsmin}) and
(\ref{upsilonstwoconstraintsmin}),  that could  be probably
measured at LHC. As to the $\psi$ family, only the 70 percentage
reduction of its contribution to the triangle diagram amplitude,
$x= -0.7$, leads to similar results: according to Eqs.
(\ref{psifamilytwoconstraintsmin}) and
(\ref{psistwoconstraintsmin}) $\Sigma_\psi BR(Z\to\gamma \psi) =
2.9\times 10^{-6}$  and $BR(Z\to\gamma J/\psi(1S))= 10^{-6}$, that
also could  be probably measured at LHC.

   The angular distributions expected in the center-of-mass system of the $q\bar q\to  Z
\to\gamma V$ and $e^+ e^-\to Z \to\gamma V$ reactions follows from
Eqs. (\ref{Ztogg*1}) and (\ref{Ztogg*2}).
\begin{equation}
\label{e+e-orqbarqgammaVangledistribution}
 W(\theta) =
\frac{3}{8}\cdot \frac{1 + \cos^2 \theta + ( 2m_q^2/M^2 )\sin^2
\theta}{1 + m_V^2/M^2}\approx \frac{3}{8}(1 + \cos^2 \theta ),
\end{equation}
where $\theta$ is the angle between the $\gamma$ quantum momentum
and the beam axis. For more details, see  the Appendix.

 I is indebted to
Pavel Murrat, whose interest and numerous discussions stimulated
the writing of this text.

This work was supported in part by the RFFI Grant No. 10-02-00016
from the Russian Foundation for Basic Research.

\begin{flushright}
{\bf Appendix }
\end{flushright}
\begin{center}
 {\bf Angle distributions}
\end{center}
  The angular distributions expected in the  $Z\to \gamma V$ decays in
  the rest frame of the $Z$ boson follow from Eqs.(\ref{Ztogg*1}) and
(\ref{Ztogg*2}).

If not to be interested in the photon and $V$ meson polarizations
  from Eqs. (\ref{Ztogg*1}) and (\ref{Ztogg*2}) it
is received, neglecting  members  $\sim (m_V/M_Z) ^2$,
\begin{equation}
\label{We(Z)n}
 W\Bigl (\mbox{\bf e($Z$)}\,,\mbox{\bf n}\Bigr )=(3/4) \biggl
(\Bigl (\mbox{\bf e($Z$)}^*\cdot\mbox{\bf e($Z$)}\Bigr ) - \Bigl
(\mbox{\bf n}\cdot\mbox{\bf e($Z$)}^*\Bigr )\Bigl (\mbox{\bf
n}\cdot\mbox{\bf e($Z$)}\Bigr )\biggr)
\end{equation}
or
\begin{eqnarray}
\label{thetadistributionswithZpolarizations} && W(S_z=1\,,\ \theta
)=W(S_Z=-1\,,\ \theta)=(3/8)(1+ cos^2\theta)\,,\\[6pt] &&
W(S_z=0\,,\ \theta)= (3/4)\sin^2\theta\,,
\end{eqnarray}
where $S_z$ is the $z$ component of the $Z$ boson spin in its rest
frame, $\theta$ is the angle between the  $\gamma$ quantum
momentum and the $z$ axis in the $Z$ boson rest frame.

If  to be interested in polarization of the photon only from
Eqs. (\ref{Ztogg*1}) and (\ref{Ztogg*2}) it is received,
neglecting members $\sim (m_V/M_Z) ^2$,
\begin{equation}
\label{We(Z)e(gamma)n} W\Bigl (\mbox{\bf e($Z$)}\,,\mbox{\bf
n}\,,\mbox{\bf e($\gamma$)}\Bigr )=(3/4)\biggl (\mbox{\bf
n}\cdot\Bigl [\mbox{\bf e($\gamma$)}\times\mbox{\bf e($Z$)}\Bigr
]\biggr )\biggl (\mbox{\bf n}\cdot\Bigl [\mbox{\bf
e($\gamma$)}\times\mbox{\bf e($Z$)}\Bigr ]\biggr )^*
\end{equation}
or
\begin{eqnarray}
\label{thetadistributionswithZphotopolarizations} && W(S_z=1\,,\
S_\gamma=+1\,,\ \theta)= W(S_z=-1\,,\ S_\gamma= -1\,,\
\theta)=(3/16) (1+\cos\theta)^2\,,\\[6pt] && W(S_z=1\,,\ S_\gamma
=-1\,,\ \theta )=W(S_z=-1\,,\ S_\gamma=+1\,,\ \theta
)=(3/16)(1-\cos\theta)^2\,,\\[6pt] && W(S_z=0\,,\ S_\gamma=+1\,,\
\theta )=W(S_z=0\,,\ S_\gamma=-1\,,\ \theta)=(3/8)\sin^2\theta\,,
 \end{eqnarray}
where $S_\gamma$ is the photon helicity.

 Note that $Z$
boson with $S_z=0$ is not produced if  the $z$ axis is the axis of
the $e^+e^-$ or $q\bar q$ beams in their center-of-mass system.
This results in Eqs. (\ref{e+e-orqbarqgammaVangledistribution}),
(\ref{e+e-Sgamma}), (\ref{ubaruSgamma}), and (\ref{dbardSgamma}).

In that event,  the angular distributions expected in the
center-of-mass system of  the  $e^+ e^-\to Z \to\gamma V$
 and $q\bar q\to  Z \to\gamma V$  reactions, $W^{e^+e^-}_{S_\gamma}(\theta)$ and $W^{q\bar q}_{S_\gamma}(\theta)$
 respectively,
are
\begin{equation}
\label{e+e-Sgamma}
 W^{e^+e^-}_{S_\gamma=\pm
 1}(\theta)=\frac{3}{16N_e}\left
 [(1/2-\xi)^2(1\mp\cos\theta)^2+\xi^2(1\pm\cos\theta)^2\right
 ]\,,
\end{equation}
where $N_e=(1/2-\xi)^2+\xi^2$, $\xi=\sin^2\Theta_W=0.23$
\cite{pdg-10}, the $z$ axis is put in the electron momentum
direction,
\begin{equation}
\label{ubaruSgamma}
 W^{u\bar u}_{S_\gamma=\pm
 1}(\theta)=\frac{3}{16N_u}\left
 [(1/2-e_u\,\xi)^2(1\mp\cos\theta)^2+e^2_u\,\xi^2(1\pm\cos\theta)^2\right
 ]\,,
\end{equation}
where $N_u=(1/2-e_u\,\xi)^2+e^2_u\,\xi^2$, $e_u=2/3$ ; the $z$
axis is put in the $u$ quark momentum direction, and
\begin{equation}
\label{dbardSgamma}
 W^{d\bar d}_{S_\gamma=\pm
 1}(\theta)=\frac{3}{16N_d}\left
 [(1/2-e_d\,\xi)^2(1\mp\cos\theta)^2+e^2_d\,\xi^2(1\pm\cos\theta)^2\right
 ]\,,
\end{equation}
where $N_d=(1/2-e_d\,\xi)^2+e^2_d\,\xi^2$, $e_d=-\ 1/3$ ; the $z$
axis is put in the $d$ quark momentum direction.

\end{document}